\documentstyle[preprint,aps,12pt,epsfig,floats]{revtex}
\tightenlines

\begin{document}
\newcommand{\beq}{\begin{equation}}
\newcommand{\eeq}{\end{equation}}
\newcommand{\beqa}{\begin{eqnarray}}
\newcommand{\eeqa}{\end{eqnarray}}
\newcommand{\fr}{\frac}
\newcommand{\sect}[1]{\section{#1}\setcounter{equation}{0}}
\renewcommand{\theequation}{\arabic{section}.\arabic{equation}}
\newcommand{\rf}[1]{\ref{#1}}
%\draft
\preprint{INJE-TP-03-05, hep-th/0304231}

\title{ Difference between AdS and dS spaces : wave equation approach}

\author{ Y. S. Myung\footnote{Email-address :
ysmyung@physics.inje.ac.kr} and N. J. Kim\footnote{Email-address :
dtpnjk@ijnc.inje.ac.kr}}
\address{Relativity Research Center and School of Computer Aided Science,
Inje University, Gimhae 621-749, Korea}

\maketitle
\vspace{5mm}

\begin{abstract}
We study the wave equation for  a  massive scalar field
 in three-dimensional AdS-black hole and
 dS (de Sitter) spaces to find what is the difference and similarity
between two spaces. Here the AdS-black hole is provided by the
$J=0$ BTZ black hole. To investigate  its event (cosmological)
horizons, we compute the absorption cross section, quasinormal
modes, and study the AdS(dS)/CFT correspondences. Although there
remains an unclear point in defining the ingoing flux near
infinity of the BTZ black hole, quasinormal modes are obtained and
the AdS/CFT correspondence is confirmed.  However, we do not find
quasinormal modes and thus do not confirm the assumed dS/CFT
correspondence. This difference between AdS-black hole and dS
spaces is very interesting, because their global structures are
similar to each other.
\end{abstract}
%\vfill
%Compiled at \today : \number \time.

\newpage

\sect{Introduction} Recently an accelerating universe has proposed
to be a way to interpret the astronomical data of
supernova\cite{Per,CADS,Gar}. The inflation is employed to solve
the cosmological flatness and horizon puzzles arisen in the
standard cosmology. Combining this observation with the need of
inflation
 leads to that our universe approaches de Sitter
geometries in both the infinite past and the infinite
future\cite{Wit,HKS,FKMP}. Hence it is  important to study the
nature of de Sitter (dS) space and the assumed dS/CFT
correspondence\cite{BOU,cons,STR}.
 However,
there exist  some difficulties in studying de Sitter space with a
positive cosmological constant $\Lambda_{dS}>0$. i) There is no
spatial (timelike) infinity and global timelike Killing vector.
Thus it is not easy to define conserved quantities including mass,
charge and angular momentum appeared in asymptotically  dS space.
ii) The dS solution is absent from string theories and thus we do
not get a definite example to test the assumed dS/CFT
correspondence. iii) It is hard to define  the $S$-matrix because
of the presence of the cosmological horizon. For the AdS-black
hole
 with a negative cosmological constant $\Lambda_{AdS}<0$
 \footnote{Here we wish to distinguish
 the pure AdS space and the AdS-black hole ($J=0$ BTZ black hole) space.
 The latter is obtained by identifying
 points in the pure AdS space.
 Also  the AdS-black hole space does not have  globally timelike Killing
vector but the AdS space has it. In this work we are interested
mainly in the AdS-black hole.}, all of three difficulties
mentioned in dS space seem to be resolved even though lacking for
a globally timelike Killing vector\cite{Wit}. There exists a
spatial (timelike) infinity and hence the region outside the event
horizon is noncompact. Many AdS solutions are arisen from string
theory or $M$-theory. There is no notion of an $S$-matrix in
asymptotically AdS space, but one has the correlation functions of
the boundary CFT. The correlators of the boundary CFT may provide
the $S$-matrix elements of the $\Lambda \to 0$ theory.

We remind the reader that the cosmological horizon in dS space is
very similar to the event horizon of the balck hole in the sense
that one can define its thermodynamic quantities of a temperature
and an entropy using the same way as was done for the black
hole\cite{GHaw}. Two important quantities in studying the black
hole are the Bekenstein-Hawking entropy and the absorption cross
section (greybody factor). The former relates to the intrinsic
property of the black hole itself, while the latter relates to the
effect of spacetime curvature on the propagation of the perturbed
wave. In other words, the computation of the absorption cross
section is based on the solution to the  wave equation. Explicitly
the greybody factor for the black hole arises as scattering of a
wave off the gravitational potential surrounding the
horizon\cite{grey1}. The low-energy $s$-wave greybody factor for a
massless scalar wave has a universality such that it is equal to
the area of the horizon for all spherically symmetric
 black holes\cite{grey2}. Also the greybody factor measures the Hawking radiation
 in a semiclassical way.    The entropy for the cosmological horizon was
 discussed in literature\cite{entropy} and the absorption of a
 scalar wave in the cosmological horizon was investigated in
 \cite{deabs}.

 Here  we focus on
  the consequences of the wave equation approach in
 both AdS-black hole and dS spaces. In addition to the greybody factor, these
 include quasinormal modes\cite{Chan}, and AdS(dS)/CFT correspondences\cite{Malda}.
The quasinormal modes of a  scalar field were studied in the
background of black holes. The associated complex frequencies
describe the time-decay of the scalar perturbation in the black
hole background. The radiation associated with these modes is
expected to be seen with the gravitational wave detectors in the
near future. Most fields propagating in the pure AdS space can be
expanded in ordinary real normal modes\cite{HH}. That is, the
negative cosmological constant provides an effective confining
(bounded) box and solutions exist with discrete real frequencies.
However, if a black hole is present in AdS space, the fields fall
into the AdS-black hole and decay continuously. In this case
frequencies become complex, showing characteristic of the black
hole which describe the decay of the perturbation outside the
event horizon.

Also the quasinormal modes of the  AdS-black hole have a direct
interpretation in terms of the dual CFT. Using the AdS/CFT
correspondence, a black hole in AdS space corresponds to a thermal
state in the CFT approximately. Perturbing the black hole by a
bulk scalar field corresponds to perturbing the thermal state in
the CFT by making use of the corresponding operator ${\cal O}$.
The timescale for the decay of a  scalar is given by the imaginary
part ($\omega_I$) of  quasinormal frequencies. The decay of scalar
perturbation describes the return to thermal equilibrium in the
CFT. So we can obtain  a prediction for the thermalization
timescale in the two-dimensional CFT. In the three-dimensional
AdS-black hole (BTZ black hole), there is a precise quantitative
agreement between  quasinormal frequencies and  location of the
poles of the Fourier transform for the retarded correlation
function of ${\cal O}$ : ${\cal D}^{ret}(x,x')=i \theta
(t-t')<[{\cal O}(x),{\cal O}(x')]>_T$\cite{Birm3}. The set of
poles characterizes the decay of  thermal perturbation on the CFT
side. We call  this the thermal AdS/CFT correspondence. As a
result, the thermal  AdS/CFT correspondence is well established
beyond our expectation. We study  this connection to understand
the cosmological horizon in dS space and to test whether or not
the thermal dS/CFT correspondence can be realized.

In this work we investigate the consequences of  the wave equation
approach for  a massive scalar in the background of  both
 the AdS-black hole and dS spaces. As a toy model of the  AdS space
with the event horizon,  the $J=0$ BTZ black hole is
introduced\cite{CLe}. The BTZ black hole is a rotating black hole
in three dimensions, as is shown in its other name of Kerr-AdS
black hole. This is a counterpart of the Kerr-dS solution.
However,  these rotating solutions have the singularity at $r=0$.
In order to avoid the singularity,  we consider the non-rotating
($J=0$) BTZ black hole with the event horizon and dS space with
the cosmological horizon only.

At this stage  we wish to point out a  difference between the
cosmological horizon in de Sitter space and the event horizon in
the Schwarzschild black hole. The cosmological horizon is usually
assumed to be in thermal equilibrium with perturbations\cite{DDO}.
This implies that the cosmological horizon not only absobs
radiation, but also emits radiation previously emitted by itself
at the same rate, keeping the curvature radius $l$ fixed. The two
of black hole and heat bath will be in thermal equilibrium if the
box is bounded because the black hole (radiation in box) have a
negative (positive) specific heat. As an example,  the eternal
black hole in AdS space (AdS-Schwarzschild black hole in four and
higher dimensions)\cite{Mal} was introduced because AdS space is
considered as an effective confining box.
 If the box is unbounded, the black hole evaporates
completely  as the Schwarzschild black hole does. In this sense
the de Sitter cosmological  horizon is similar to the event
horizon of AdS-black hole\cite{DKS}.
 The organization of this paper is as follows. In section II we
briefly review two examples of quantum mechanics. Section III is
devoted to studying  the consequences of the  wave equation
approach in the AdS-black hole space. We study the dS wave
equation along the AdS-black hole in section IV. We summarize our
results in section V:
 what is the difference and similarity between AdS-black hole and dS spaces in the wave equation
 approach.

\sect{Two examples in quantum mechanics}

In order to obtain the absorption cross section, we have  to
calculate the outgoing/ingoing fluxes in the AdS-black hole and dS
space. However, at the timelike infinity of the AdS-black hole
space, there remains some ambiguities to define the
outgoing/ingoing fluxes. To estimate it clearly, let us review the
flux computation in the potential barrier in quantum mechanics.
Also there exists some problems in computation
 of the  flux in dS space. For this purpose,   we introduce the
 wave propagation under the potential step with $0<E<V_0$.
 This gives rise to the classical picture of
what  goes on :  the total reflection occurs  due to the potential
step.

\subsection{ Plane wave under potential step}
Let us start with a wave propagation under a potential step with
height $V_0>E$ shown in Fig.1.
\begin{figure}[b!]
\begin{center}
\epsfig{file=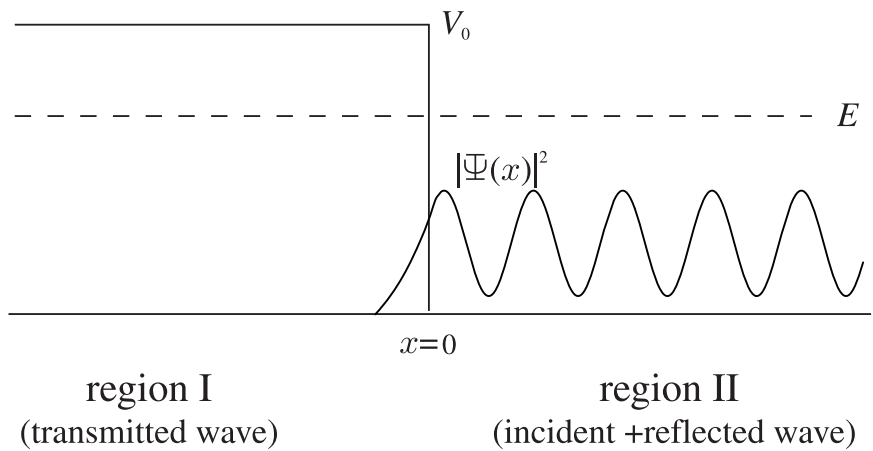,width=0.7\textwidth}\caption{The upper
portion of the figure shows the potential step $V(x)$. The total
energy $E$ is indicated by the dashed line. The lower part shows
the absolute square of the wave function $|\Psi_{\text{I}}(x)|$,
and $|\Psi_{\text{II}}(x)|$. In region II, we have a standing wave
pattern, which arises because the reflected wave interferes with
the incident wave. Also the wave penetrates into the classically
forbidden region I.}
\end{center}
\end{figure}
Here we introduce the Schr\"{o}dinger equation with $\hbar=1,
m=1/2$ as
\begin{equation}
\label{1eq2} -\frac{d^2 \Psi(x)}{dx^2}+V(x)\Psi(x)=E\Psi(x).
\end{equation}
The solution to the Schr\"{o}dinger equation is given by

\begin{eqnarray}
\label{2eq2}
\Psi_{\text{I}} (x)=Te^{qx},~~~~~~~~~~~~~~~~\text{region I} \\
\Psi_{\text{II}} (x)=e^{-ikx}+Re^{ikx},~~~~\text{region II}
\end{eqnarray}
with $k=\sqrt{E},\,q=\sqrt{V_0 -E}$.
Considering the continuity of
$\Psi(x)$ and $\Psi'(x)$ at $x=0$ leads to

\begin{eqnarray}
\label{4eq2}
&&1+R=T, \\
&&ik(1-R)=-qT.
\end{eqnarray}
The above relations take the form
\begin{eqnarray}
\label{6eq2}
&&R=\frac{1-i\tan\theta}{1+i\tan\theta}=e^{-2i\theta} \Big(~|R|^2 =1\Big), \\
&&T=2e^{-i\theta} \cos\theta
\end{eqnarray}
with $\tan\theta =q/k$. In region I the wave function is given by
\begin{equation}
\label{8eq2}
 \Psi_{\text{I}}(x)=2e^{-i\theta}\cos\theta
e^{qx},~~|\Psi_{\text{I}}(x)|^2=4\cos^2 \theta e^{2qx}.
\end{equation}
and in region II, it takes
\begin{equation}
\label{9eq2}
\Psi_{\text{II}}(x)=2e^{-i\theta}\cos(kx-\theta),~~|\Psi_{\text{II}}(x)|^2=4\cos^2
(kx-\theta).
\end{equation}
 Since the density of incident wave is
unity, its flux (${\mathcal F}_{\text{II\emph{inc}}}$) is equal to
 $-2\sqrt{E} <0$. The reflected flux ${\mathcal
F}_{\text{II\emph{ref}}}$ is given by $2\sqrt{E}>0$ and thus there
is no net flux in region II:
${\mathcal{F}}_{\text{II\emph{inc}}}+{\mathcal{F}}_{\text{II\emph{ref}}}=0$.
According to the flux conservation, we expect that there is no
flux in region I. As a check, one finds that
${\mathcal{F}}_{\text{I}}=\frac{1}{i} [ \Psi
^{*}_{\text{I}}(\Psi_{\text{I}})'- \Psi_{\text{I}}(\Psi
^{*}_{\text{I}})']=0$. Even though the probability density of
finding a particle between $x$ and $x+dx(x<0)$ is not zero, its
flux is zero. This means that the quantum mechanical picture
reduces to the classical picture of the  total reflection. A plane
wave moving under a potential step of height ($V_0>E$) corresponds
to a toy model for the total reflection with $|R|^2=1$ and
$~R=e^{-2i\theta}$ (non-zero phase). A similar situation occurs in
a scalar wave propagating under de Sitter space.

\subsection{Plane wave under potential barrier}
In this section we clarify the flux ambiguity by considering a
plane wave propagation under the potential barrier shown in Fig.2.
%\vspace{3mm}
\begin{figure}
\begin{center}
\epsfig{file=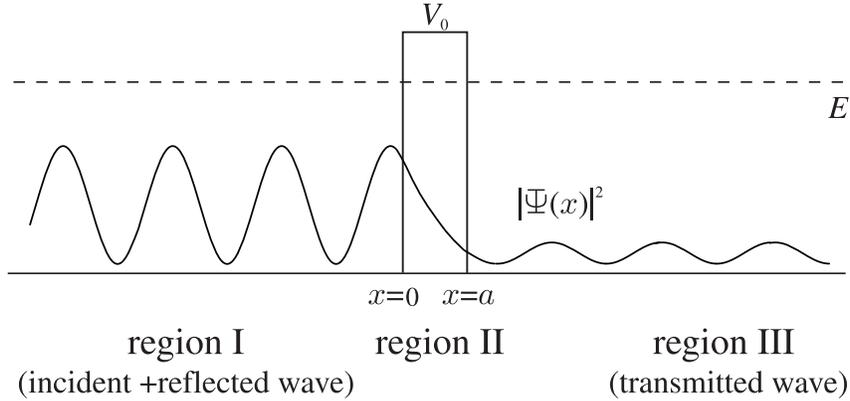,width=0.7\textwidth} \caption{The upper
portion of the figure shows the potential barrier ($V_0>E$) with
the total energy ($E$) denoted by the dashed line. The lower
portion shows the absolute square of the wave function:
$|\Psi_{\text{I}}(x)|^2,~|\Psi_{\text{II}}(x)|^2,~|\Psi_{\text{III}}(x)|^2$.
We note the transmitted wave in region III and the exponentially
decreasing function inside the barrier. In region I, we have an
imperfect standing wave pattern. Because of the conservation of
the flux in whole region, one finds ${\mathcal{F}} \neq 0$
anywhere.}
\end{center}
\end{figure}
%\vspace{10mm}
The solution to the Schr\"{o}dinger equation leads to
\begin{eqnarray}
\label{10eq2}
\Psi_{\text{I}}(x)= e^{ikx}+R e^{-ikx},~~~\text{region I} \\
\Psi_{\text{II}}(x)= Ae^{qx}+B e^{-qx},~~~\text{region II} \\
\Psi_{\text{III}}(x)= T e^{ikx},~~~\text{region III}
\end{eqnarray}
with $k=\sqrt{E},~ q=\sqrt{V_0 -E}$. Imposing the boundary
condition on the wave function and its derivative at $x=0$ and
$x=a$, one finds four relations:
\begin{eqnarray}
\label{13eq2}
&&1+R=A+B, \\
&&1-R=-i\rho(A-B), \\
&& Ae^{\theta}+B e^{-\theta}=Te^{i\delta}, \\
&& Ae^{\theta}-B e^{-\theta}=i\frac{T}{\rho}e^{i\delta},
\end{eqnarray}
where $\rho=q/k,~\theta=qa,~\delta=ka$. From the above relations,
we obtain  relevant quantities
\begin{eqnarray}
\label{17eq2}
&&A=\frac{T}{2}e^{(i\delta -\theta)} (1+\frac{i}{\rho}), \\
&&B=\frac{T}{2}e^{(i\delta +\theta)} (1-\frac{i}{\rho}),
\end{eqnarray}
where
 \beq
 \label{19eq2}
  T=\fr{e^{i\delta}}{\cosh \theta
  +\fr{i}{2}(\rho-\fr{1}{\rho})\sinh\theta}.
 \eeq
 Here we focus on  computing the flux at region II.
 At the first sight, we might have thought that ${\mathcal
F}_{\text{II}}=0$, since the wave function $\Psi_{\text{II}}(x)$
does not take a form of travelling wave.
 According to the flux conservation, we expect that
${\mathcal F}_{\text{I}}={\mathcal F}_{\text{II}}={\mathcal
F}_{\text{III}}$, otherwise waves  would be created or destroyed.
Actually one has a non-zero flux\cite{quan}
\begin{equation}
\label{20eq2} {\mathcal{F}}_{\text{II}} = 4q \,\text{Im}[AB^{*}].
\end{equation}
This shows that if the coefficient $A$ or $B$ is zero, or if they
are both real (in general, if they are complex but have the same
phase), the flux ${\mathcal{F}}_{\text{II}}$ is indeed zero. In
the previous subsection, the coefficient corresponding to $A$ is
zero and thus its flux is zero. In general potential barrier
problem, neither $A$ nor $B$ is zero, and their relative phase
gives a non-zero flux. However, we  note that it is not easy to
determine whether the value of the flux $\mathcal{F}_{\text{II}}$
in Eq.(\ref{20eq2}) is positive or negative. A similar situation
to this case occurs in the scalar wave propagation near infinity
in the BTZ black hole background.

\sect{ wave equation in AdS-black hole space}

\subsection{Wave equation}
We start with the wave propagation for a massive scalar field
with mass $m$

\beq (\nabla_{BTZ}^2 -m^2) \Phi_{BTZ}=0 \label{1eq3}
 \eeq
in the background of  the $J=0$ BTZ black hole  whose line element
is given by\cite{CLe}
 \beq
 ds_{BTZ}^2=-\Big(-M+
\fr{r^2}{l^2} \Big) dt^2 + \Big(-M+ \fr{r^2}{l^2} \Big)^{-1}dr^2
+r^2 d\phi^2,
 \label{2eq3}
 \eeq
where $M$ is the mass of the black hole and
  $l$ is the curvature radius
of AdS space. Hereafter we set $M=l=1$ for simplicity unless
otherwise stated. The above metric is singular at the event
horizon of $r=r_{EH} =1$, which divides space into four regions.
Its global structure (Penrose diagram) is shown in Fig.3.
\begin{figure}[b!]
\begin{center}
\epsfig{file=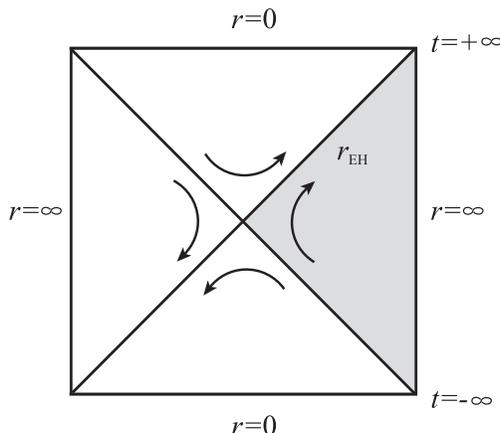,width=0.4\textwidth} \caption{The global
structure of the $J=0$ BTZ black hole spacetime. The causally
connected region (shaded region : SR)  is noncompact. Hence we
expect that the AdS/CFT correspondence which is usually defined at
$r=\infty$ is  established in the SR.}
\end{center}
\end{figure}
There are two regions with $0\le r < r_{EH}$ which correspond to
the interior region of the $J=0$ BTZ black hole. Two regions with
$r_{EH}\le r \le\infty$ correspond to the exterior of the black
hole. A timelike infinity of $r=\infty$ appears but there does not
exist any spacelike infinity in AdS-black hole space. A timelike
Killing vector $\fr{\partial}{\partial t}$ is future-directed only
in the shaded diamond. This means that there is no globally
defined timelike Killing vector in AdS-black hole space. To obtain
the greybody factor, we have to obtain a definite wave propagation
as time evolves. Hence  we confine ourselves to the shaded region
(SR). This means that our working space is noncompact, in contrast
to the case of dS space. A scalar with mass $m^2\ge-1$ including
the tachyon with $m^2=-3/4$ may be allowed for  AdS space. A
massless scalar with $m^2=0$ is special and it would be treated
separately. Hence we are interested in the massive scalar
propagation with $m^2\ge 0$.

Assuming a mode solution \beq \Phi_{BTZ}(r,t,\phi)=f_{\ell}(r)
e^{-i \omega t} e^{i \ell\phi}, \label{3eq3} \eeq the radial part
of the wave equation is \beq (r^2 -1)f_{\ell}''(r) - \Big( \fr{1}
{r} -3r \Big) f_{\ell}'(r) + \Big( \fr{\omega^2}{r^2 -1}
-\fr{\ell^2}{r^2} -m^2 \Big) f_{\ell}(r)=0, \label{4eq3} \eeq
where the prime ($'$) denotes the differentiation with respect to
its argument. Hereafter one should distinguish $\ell$ of the
angular momentum number from $l$ of the curvature radius of AdS
space.

\subsection{Potential analysis}

We observe from Eq.(\ref{4eq3}) that
 it is not easy to see how a scalar wave propagates in the exterior of the $J=0$ BTZ black hole.
  In order to do that, we must transform the wave equation
into the Schr\"odinger-like equation by introducing a tortoise
coordinate $r^*$\cite{CMann,ML}.  Then we can get asymptotic
waveforms in regions of $r^*=-\infty$ and $r^*=0$ through a
potential analysis. For our purpose, we introduce  \beq r^*=
\fr{1}{2} \ln\Big[\fr{r-1}{r+1}\Big],~~ r=-\coth r_* \label{5eq3}
\eeq to transform Eq.(\ref{4eq3}) into the Schr\"odinger-like
equation with the energy $E=\omega^2$
 \beq -\fr{d^2}{d r^{*2}} f_{\ell} + V_{BTZ}
(r)f_{\ell}=  E f_{\ell}. \label{6eq3}
 \eeq
Here  the BTZ potential is given by\cite{CLe}
 \beq V_{BTZ}(r)= (r^2 -1) \Big[ \fr{3}{4}+m^2
+\fr{\ell^2}{r^2}+\fr{1}{4r^2} \Big]. \label{7eq3}
 \eeq
 From Eq.(\ref{5eq3}) we confirm that $r^*$ is a
tortoise coordinate such that $r^* \to -\infty (r \to 1)$, whereas
$r^* \to 0 (r \to \infty)$. We  express the BTZ potential as a
function of $r^*$
 \beq V_{BTZ}(r^*) = \Big[
\Big(\fr{3}{4}+m^2\Big)\coth^2 r^* -m^2+\ell^2-\fr{1}{2}-(\ell^2
+\fr{1}{4})\tanh^2 r^* \Big]. \label{8eq3} \eeq
 We observe that for $m^2>0$,
  $V_{BTZ}(r^*)$ decreases exponentially to zero as one approaches the event horizon ($r^* \to
 -\infty,r \to r_{EH}=1$), while it goes  infinity as one approaches
infinity($r^* \to 0,r \to \infty$). A typical form of $V_{BTZ}
(r^* )$ is shown in Fig.4. \vspace{5mm}
\begin{figure}
\begin{center}
\epsfig{file=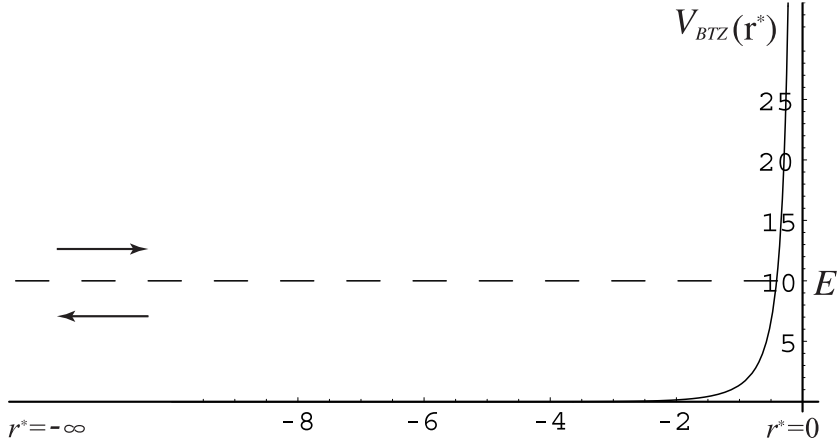,width=0.7\textwidth} \caption{The shape of
$V_{BTZ}(r^* )$ with $m^2 =1,~\ell^2 =0,~ E=10$. The ingoing wave
($\leftarrow$) and the outgoing wave ($\to$) are developed near
the event horizon of $r^* =-\infty$. $r^* =0$ corresponds to
infinity of $r=\infty$. At this point, it is subtle to define the
ingoing flux.}
\end{center}
\end{figure}
We thus conjecture that travelling waves appear near the event
horizon of $r^*=-\infty$. But  it is not easy to develop a genuine
travelling wave at infinity.  One finds an approximate equation
near $r^*=0$

\beq \fr{d^2f_{\ell,0}}{d r^{*2}}  -\big(
\fr{3}{4}+m^2\big)\fr{f_{\ell,0}}{r^{*2}}=0. \label{9eq3} \eeq
This gives us a solution

\beq f_{\ell,0}(r^*) =A_{BTZ} ~e^{( 1+\sqrt{1+m^2} )\ln r^*} +
B_{BTZ} ~e^{( 1-\sqrt{1+m^2} )\ln r^*}. \label{10eq3} \eeq For
$m^2
>0$, the first term corresponds to a normalizable mode at
$r=\infty~(r^*=0)$, while the second term is a non-normalizable
mode.

On the other hand, near the event horizon $r_{EH}=1(r^*=-\infty)$
one obtains an approximate  equation because $V_{BTZ} (r^* )\to 0$
as $r^* =-\infty$

\beq \fr{d^2}{d r^{*2}} f_{\ell,-\infty} +\omega^2
f_{\ell,-\infty}=0 \label{11eq3} \eeq which gives us   a
travelling wave solution

\beq f_{\ell,-\infty}(r^*) =C_{BTZ}~ e^{-i \omega r^*} + D_{BTZ}
~e^{i \omega r^*}, \label{12eq3} \eeq where considering
$e^{-i\omega t}$, the first term is the ingoing wave
($\leftarrow$), and the second is the outgoing wave
($\rightarrow$).

\subsection{Flux calculation}
Up to now we obtain the approximate solutions near $r^*=-\infty,
0$. In order to obtain the  solution which is valid for whole
region outside the black hole, we  solve equation (\ref{4eq3})
explicitly. We wish to transform it into a hypergeometric equation
using $z=(r^2-1)/r^2$. Our working space  remains unchanged as $0
\le z \le 1$ covering the exterior of the BTZ black hole. This
equation takes a form

\beq z(1-z)f_{\ell}''(z) +(1-z) f_{\ell}'(z) + \fr{1}{4} \Big(
\fr{\omega^2}{z} -\fr{\ell^2}{1-z} -m^2 \Big) f_{\ell}(z)=0.
\label{13eq3} \eeq Here one finds two poles at $z=0,1
(r=1,\infty)$ and so makes a further transformation  to cancel
these by choosing an ingoing solution at $z=0(r=r_{EH})$

\beq f(z)=z^\alpha (1-z)^\beta w(z),~~ \alpha =-
\fr{i\omega}{2},~~\beta= \fr{1}{2}(1- \sqrt{1+m^2}). \label{14eq3}
\eeq
  Then we obtain the hypergeometric equation \beq
z(1-z)w''(z) + [c-(a+b+1)z] w'(z) -a b~ w(z) =0, \label{15eq3}
\eeq where $a,b$ and $c$ are given by

\beq
 a= \fr{1}{2} (i \ell - i \omega +1-\sqrt{1+m^2}),~~ b= \fr{1}{2} ( -i\ell - i \omega
 +1-\sqrt{1+m^2}),~~ c= 1+-i\omega.
 \label{16eq3}
 \eeq
The ingoing solution near $z=0$ to Eq.(\ref{13eq3}) is chosen
 as\footnote{In the neighborhood of the event horizon, the two linearly independent
 solution to Eq.(\ref{15eq3}) are given by $w(z)=F(a,b,c;z)$ and
 $w(z)=z^{1-c}F(a-c+1,b-c+1,2-c,;z)$\cite{AS}.
 But the latter corresponds to the outgoing wave and thus we
 choose the former for our purpose.}
 \beq
 f(z)
 =C_{BTZ} ~z^{\alpha}(1-z)^{\beta} F(a,b,c;z)
 \label{17eq3}
 \eeq
which is in accordance with Eq.(\ref{12eq3}) with $D_{BTZ}=0$. Now
we are in a position to calculate an  ingoing flux\footnote{In
this work, one defines ``ingoing flux" as negative whereas
``outgoing flux" is defined to be positive.} at
$z=0(r=1,r^*=-\infty)$ which shows

\beq {\cal F}_{\text{\emph{in}}}(z=0)= 2\fr{2 \pi}{i}
[f^*z\partial_z f-f z\partial_z f^*]|_{z=0}=-4\pi\omega
|C_{BTZ}|^2. \label{18eq3} \eeq  To obtain the flux at infinity
$z=1(r=\infty,r^*=0)$, we  use a formula:

\beqa \label{19eq3}
&&F(a,b,c;z)=
\fr{\Gamma(c)\Gamma(c-a-b)}{\Gamma(c-a)\Gamma(c-b)}
F(a,b,a+b-c+1;1-z)\\ \nonumber && ~~+
\fr{\Gamma(c)\Gamma(a+b-c)}{\Gamma(a)\Gamma(b)}(1-z)^{c-a-b}
F(c-a,c-b,-a-b+c+1;1-z)  \eeqa
 with
  \beqa
  \label{20eq3}
   c-a=\fr{1}{2}[-i\ell-i\omega+1+\sqrt{1+m^2}]\neq a^*,  \nonumber\\
   c-b=\fr{1}{2}[i\ell-i\omega+1+\sqrt{1+m^2}] \neq b^*, \nonumber \\
   c-a-b=\sqrt{1+m^2}\neq(a+b-c)^*.
 \eeqa
We note here that by comparing Eq.(\ref{16eq3}) with
Eq.(\ref{20eq3}), $c-a-b,~c-a,~c-b$ are not complex conjugates of
$a+b-c,~a,~b$, respectively. This is because we consider only the
case of $m^2>0$. Using $1-z \approx r^{*2}(\approx 1/r^2)$ near
infinity, one finds from Eqs.(\ref{17eq3}) and (\ref{19eq3}) the
following form:

\beqa \label{21eq3}
 f_{0 \to 1} \equiv f_{nor} + f_{non}
 =H_{nor} e^{(1+\sqrt{1+m^2})\ln r^*} + H_{non}e^{(1-\sqrt{1+m^2})\ln
 r^*}\\
 =H_{nor}e^{-(1+\sqrt{1+m^2})\ln r} +H_{non} e^{(-1+\sqrt{1+m^2})\ln r},
 \eeqa
 where the explicit forms of normalizable  and nonnormalizable amplitudes are given by
 \beqa
 \label{23eq3}
 &&H_{nor}=C_{BTZ}E_{nor},~~ E_{nor}=
\fr{\Gamma(c)\Gamma(a+b-c)} {\Gamma(a) \Gamma(b)}, \\
&&H_{non}=C_{BTZ}E_{non} ,~~ E_{non}= \fr{\Gamma(c)\Gamma(c-a-b)}
{\Gamma(c-a) \Gamma(c-b)}.
  \eeqa
Then we can match Eq.(\ref{10eq3}) with Eq.(\ref{21eq3}) near
infinity to yield

\beq A_{BTZ}=H_{nor},~~~ B_{BTZ}=H_{non}. \label{25eq3} \eeq

Finally we calculate the flux near $z=1(r=\infty)$ as

\beqa \label{26eq3} {\cal F}(z=1)= \fr{2 \pi}{i} [f_{0\to 1}^*
(r^3\partial_{r} f_{0\to1})-f_{0\to1}(r^3
\partial_{r} f_{0\to1}^*)]|_{r=\infty} \nonumber \\
= 8 \pi {\sqrt{1+m^2}} ~\text{Im} [H_{nor}^* H_{non}]
 \eeqa
which is similar to the case of potential barrier in
Eq.(\ref{20eq2}). This shows that if the coefficient $E_{nor}$ or
$E_{non}$ is zero, or if they are both real (in general, if they
are complex but have the same phase), the flux is indeed zero.  In
the present case, neither $E_{nor}$ nor $E_{non}$ is zero, and
both belong to complex. Their relative phase gives a non-zero
flux. However we have to say that it is not easy to determine
whether the value of the flux ${\cal F}(z=1)$ in Eq.(\ref{26eq3})
is positive or negative.

\subsection {Absorption cross section}
 Up to now we do not insert
the curvature radius $l$ of AdS space. The correct absorption
coefficient can be recovered when replacing $\omega(m)$ with $
\omega l(m l)$ inspired from $r \to r/l$. An absorption
coefficient by the event horizon is defined  by

 \beq {\cal A}_{BTZ} = \fr{ {\cal F}_{in}(z=0)}{{\cal F}_{in}(z=1)}
\label{27eq3}
 \eeq
 where ${\cal F}_{in}(z=0)$ is  given by
Eq.(\ref{18eq3}). However it is not easy to separate the ingoing
flux from ${\cal F}(z=1)$ in Eq.(\ref{26eq3}). Hence we may use
${\cal F}(z=1)$ instead of ${\cal F}_{in}(z=1)$ for a further
study. We follow the conventional approach to obtain the
absorption cross section in black hole physics\cite{deabs}. The
absorption cross section in three dimensions is formally defined
by

\beq
 \sigma_{abs}^{BTZ}= \fr{{\cal A}_{BTZ}}{\omega} =
-\fr{\ell}{2\sqrt{1+(ml)^2} ~\text{Im} [ E_{nor}^* E_{non}]}.
\label{28eq3} \eeq Explicitly, $E_{nor}$ and $E_{non}$ are given
by \beqa
 \label{29eq3}
  E_{nor}^*=
\fr{\Gamma(1+ i \omega l)\Gamma(-\sqrt{1+(ml)^2})} {
\Gamma[(i\omega l-i\ell +1 -\sqrt{1+(ml)^2})/2] \Gamma[(i\omega l
+i \ell +1
-\sqrt{1+(ml)^2})/2]},\\
\label{30eq3}
 E_{non}= \fr{\Gamma(1- i
\omega l)\Gamma(\sqrt{1+(ml)^2})} {\Gamma[(-i\omega l-i \ell +1
+\sqrt{1+(ml)^2})/2]\Gamma[(-i\omega l + i\ell +1
+\sqrt{1+(ml)^2})/2]}.
  \eeqa
For $(ml)^2=-1$, one finds that  $E^*_{nor}E_{non} \to
|E_{nor}|^2$ (real) and thus ${\cal F}(z=1) \to 0$ irrespective of
the factor $\sqrt{1+(ml)^2}$. For a massless scalar propagation
with $m^2=0$, one has
 \beqa
 \label{31eq3}
E_{nor}^*= \fr{\Gamma(1+ i \omega l)\Gamma(-1)} {
\Gamma[(i\omega l-i\ell)/2] \Gamma[(i\omega l +i \ell )/2]},\\
 E_{non}= \fr{\Gamma(1- i
\omega)\Gamma(1)} {\Gamma[(-i\omega l-i \ell
)/2+1]\Gamma[(-i\omega l+ i\ell)/2+1]}.
  \eeqa
Here we find that $E^*_{nor}E_{non} \to$ real and thus  ${\cal
F}(z=1) \to 0$, although there exists a divergent term of
$\Gamma(-1)=\pi/\sin[2\pi]$ in $E^*_{nor}$. We do not consider the
massless scalar propagation  because of this divergent term.
 Even for the $s(\ell=0)$-wave scalar
propagation, it is difficult to split ${\cal F}(z=1)$ into ${\cal
F}_{in}(z=1)$ and ${\cal F}_{out}(z=1).$ For the case of $m^2>0$,
one cannot find the correct ingoing flux from Eq.(\ref{26eq3}) to
compute the absorption cross section. This mainly arises from the
fact that one cannot find an asymptotically flat region which is
necessary for defining the scattering process well. In general, we
need a spatial infinity to obtain an absorption cross section in
the wave equation approach. Here ``spatial infinity" means
$r=\infty$ in asymptotically flat space, while ``spatial
(timelike) infinity" denotes $r=\infty$ in asymptotically AdS
space. We note that the previous calculations\cite{Birm1,ML} based
on the  wave equation approach in the BTZ black hole belong to the
approximate ones which are required to match with the results of
the CFT. Hence we conclude that the absorption cross section is
not derived from the bulk scalar propagation in AdS-black hole
space, but it can be obtained from the dual CFT correlation
function\cite{Birm3} as the scattering matrix does\cite{Wit}.

\subsection{Quasinormal modes}
Even though we encounter some difficulties in calculating the
absorption cross section in the BTZ black hole background, we can
calculate the quasinormal modes. In asymptotically flat space like
in the Schwarzschild black hole\cite{Chan}, these modes are
usually defined as the solutions which are purely ingoing wave
($\Phi \sim e^{-i \omega(t+ r^*)}$) at the event horizon and which
purely outgoing ($\Phi \sim e^{-i \omega(t -r^*)}$) near infinity
(see Fig.5). The initial ingoing wave from infinity is not
allowed.
\begin{figure}[b!]
\begin{center}
\epsfig{file=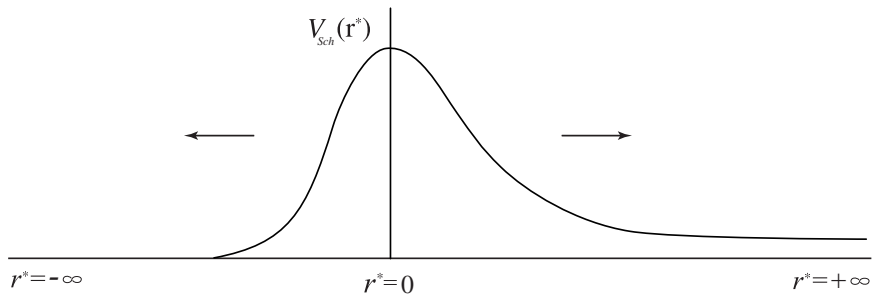,width=0.7\textwidth} \caption{Shape of
potential $V_{Sch}(r^*)$ and its quasinormal modes of a massless
scalar in the Schwarzschild black hole background. The ingoing
wave ($\leftarrow$) at the event horizon and the outgoing wave
($\rightarrow$) at infinity denote  quasinormal modes.}
\end{center}
\end{figure}

\begin{figure}[t!]
\begin{center}
\epsfig{file=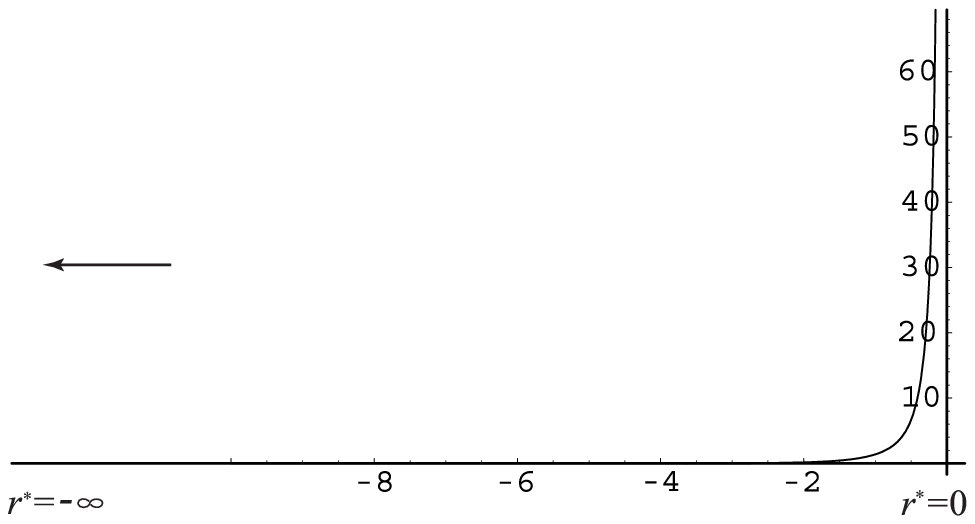,width=0.7\textwidth} \caption{The
quasinormal modes in AdS space. These are denoted by purely
ingoing modes ($\leftarrow$) at the event horizon.}
\end{center}
\end{figure}

In asymptotically  AdS space\cite{HH}, these modes are
 defined as the solutions which are purely ingoing wave
($\Phi_{BTZ} \sim e^{-i \omega(t+ r^*)}$) at the event horizon and
which vanish ($\Phi_{BTZ} \sim 0$) at infinity (see Fig.6). This
comes from the boundary condition that the wave function is zero
at infinity because its potential is infinite at infinity. Because
one cannot find any travelling wave at infinity, we can extend
this definition to the flux boundary condition in AdS-black hole
space : the ingoing flux $({\cal F}_{in}(z=0)<0)$ at the event
horizon and no flux (${\cal F}(z=1)=0$) at infinity. Then there
are two ways to obtain ${\cal F}(z=1)=0$ with the complex
frequency. First one chooses $E_{non}=0$ to get the quasinormal
modes. These impose the restriction as \beq \label{33eq3}
c-a=-n,~~c-b=-n,\eeq where $n=-1,-2,\cdots$(negative integers).
Considering Eq.(\ref{30eq3}) leads to \beqa \label{34eq3}
\omega_1= -\fr{\ell}{l} -2\fr{i}{l} \Big(n+\fr{1}{2}
+\fr{\sqrt{1+(ml)^2}}{2} \Big),\\\label{35eq3}
\omega_2=\fr{\ell}{l} -2\fr{i}{l} \Big(n+\fr{1}{2}
+\fr{\sqrt{1+(ml)^2}}{2} \Big). \eeqa Second from $E_{nor}^*=0$,
one finds $a=-n,b=-n$ which lead to \beqa \label{36eq3} \omega_3=
-\fr{\ell}{l} +2\fr{i}{l} \Big(n+\fr{1}{2}
-\fr{\sqrt{1+(ml)^2}}{2} \Big),\\\label{37eq3}
\omega_4=\fr{\ell}{l} +2\fr{i}{l} \Big(n+\fr{1}{2}
-\fr{\sqrt{1+(ml)^2}}{2} \Big). \eeqa When decomposing the
quasinormal frequencies  into  real and imaginary parts :
$\omega=\omega_R -i\omega_I$, $w_I$ should be positive for all
quasinormal frequencies because  those modes decay with time.
Hence the true quasinormal frequencies are given by
Eqs.(\ref{34eq3}) and (\ref{35eq3}) but not by Eqs.(\ref{36eq3})
and (\ref{37eq3}). We note that in calculating quasinormal modes
in AdS space, one has to use a definition of the zero-flux at
infinity. Someone uses the Dirichlet boundary condition on the
wave function at infinity which eventually leads to $E_{non}=0$. A
single term of either normalizable one or nonnormalizable one
leads to zero-flux at infinity. But requiring the Dirichlet
condition (no nonnormalizable term) at infinity leads to a correct
definition of quasinormal frequencies.

\subsection{AdS/CFT correspondence}
The complex quasinormal frequencies describe the decay of the
massive scalar perturbation in the background of AdS-black hole.
These depend on the parameters of AdS- black hole ($M,l$) as well
as the parameters of perturbed field ($m,\ell$). In terms of the
AdS/CFT correspondence, an off-equilibrium configuration in the
AdS-black hole space ($J=0$ BTZ black hole) is related to an
off-equilibrium thermal state in the boundary conformal field
theory. The timescale for the decay of scalar perturbation is
given by the imaginary part of the quasinormal frequencies
($\omega_I$). Using the AdS/CFT correspondence, one can obtain a
prediction of the timescale for return to equilibrium of its dual
CFT. Actually there is a precise agreement between quasinormal
modes on the bulk side and the location of the poles for the
retarded correlation function describing the linear response of an
perturbing operator ${\cal O}$ on the CFT side\cite{Birm3}. This
provides a new quantitative test for the thermal AdS/CFT
correspondence. Further, expressing $\omega_I \sim 2/l$ as
$\omega_I \sim \fr{r_{EH}}{\gamma} \fr{1}{r_{EH}^2}$, where
$\gamma$ is the Choptuik scaling parameter, it suggests a deeper
connection between the critical phenomena of the BTZ black hole
thermodynamics and quasinormal modes\cite{Birm2}. According to the
thermal AdS/CFT correspondence, this leads to the boundary CFT
interpretation of Choptiuk scaling (a universal scaling behavior).

\sect{ wave equation in de Sitter space}

\subsection{dS wave equation}

We start with the wave equation for a massive scalar
field\cite{BMS,deabs}\beq (\nabla_{dS}^2 -m^2) \Phi_{dS}=0
\label{1eq4} \eeq in the background of three dimensional de Sitter
space \beq ds_{dS}^2=-\Big(1- \fr{r^2}{l^2} \Big) dt^2 + \Big(1-
\fr{r^2}{l^2} \Big)^{-1}dr^2 +r^2 d\phi^2. \label{2eq4} \eeq  Here
$l^2$ is the curvature radius of de Sitter space and hereafter we
set $l=1$ for simplicity unless otherwise stated. The above metric
is singular at the cosmological horizon $r=r_{CH}=1$ which divides
space into four regions, as is shown in Fig.7.
\begin{figure}
\begin{center}
\epsfig{file=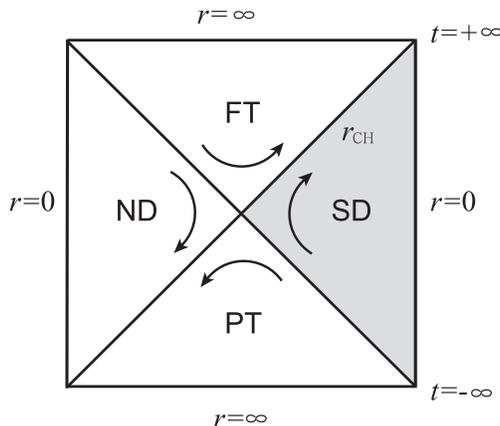,width=0.4\textwidth} \caption{The global
structure of de Sitter spacetime. This diagram looks like that of
the $J=0$ BTZ black hole   except replacing $r=0(r=\infty)$ here
by $r=\infty(r=0)$ in Fig. 3. But the difference is that the
causally connected region (shaded SD) in this figure is compact,
whereas the causally connected region (shaded region: SR) in Fig.
3 is noncompact. Hence we expect that the assumed  dS/CFT
correspondence which is usually defined in the FT or PT including
$r=\infty$ is not established in the SD where is causally
disconnected to the FT and PT.}
\end{center}
\end{figure}
There are two regions with $0\le r \le1$ which correspond to the
causal diamonds of observers at the north and south poles :
northern diamond (ND) and southern diamond (SD). An observer at
$r=0$ is surrounded by a cosmological horizon at $r=1$. Two
regions with $1 < r \le \infty$ containing the future-null
infinity ${\cal I}^+$ and past-null infinity  ${\cal I}^-$ are
called future triangle (FT) and past triangle (PT), respectively.
A timelike Killing vector $\fr{\partial}{\partial t}$ is
future-directed only in the southern diamond. To obtain the
greybody factor, we have to obtain a definite wave propagation as
time evolves. Hence  we confine ourselves to the southern diamond
(shaded region). This means that our working space  here is
compact, in contrast to the AdS-black hole.

In connection with the assumed dS/CFT correspondence, one may
classify the mass-squared $m^2 \ge 0$ into three cases\cite{STR}:
$m^2\ge1,~~0<m^2<1,~~m^2=0$. For a massive scalar with  $m^2\ge1$,
one has a non-unitary CFT. A scalar with mass $0<m^2<1$ can be
related to a unitary CFT.  A massless scalar with $m^2=0$ is
special and it would be treated separately. Here we consider only
$m^2(0<m^2<1)$ as a parameter at the beginning. Assuming a mode
solution \beq \Phi_{dS}(r,t,\phi)={\tilde f}_{\ell}(r) e^{-i
\omega t} ~e^{i\ell\phi}, \label{3eq4} \eeq Eq.(\ref{1eq4}) leads
to the differential equation for $r$\cite{BMS,ACL} \beq
(1-r^2){\tilde f_{\ell}''}(r) + \Big( \fr{1} {r} -3r \Big) {\tilde
f}_{\ell}'(r) + \Big( \fr{\omega^2}{1-r^2} -\fr{\ell^2}{r^2} -m^2
\Big) {\tilde f}_{\ell}(r)=0, \label{4eq4} \eeq where the prime
($'$) denotes the differentiation with respect to its argument.
Apparently this equation can be obtained from equation
(\ref{4eq3}) by replacing $\ell^2,m^2$ by $-\ell^2,-m^2$,
respectively. However, an important thing  is that  the working
space  is different : the working region  of  AdS space (SR) is
from $1 \le r \le \infty$, while that of dS space (SD) is $0 \le
r\le 1$. That is, there is no notion of spatial (timelike)
infinity in the SD of dS space.

\subsection{dS potential analysis}

We observe from Eq.(\ref{4eq4}) that
 it is not easy to find how a scalar wave propagates in the southern
diamond. In order to do that, we must transform the wave equation
into the Schr\"odinger-like equation using a tortoise coordinate
$r^*$\cite{deabs}.  Then we can get wave forms in asymptotic
regions of $r^*\to \pm \infty$ through a potential analysis. We
introduce $r^*=g(r)$ with $ g'(r)=1/r(1-r^2)$ to transform
Eq.(\ref{4eq4}) into the Schr\"odinger-like equation with the
energy $E=\omega^2$

\beq -\fr{d^2}{d r^{*2}} {\tilde f}_{\ell} + V_{dS}(r){\tilde
f}_{\ell}=  E {\tilde f}_{\ell} \label{5eq4} \eeq with the dS
potential \beq V_{dS}(r)= \omega^2 + r^{2}(1-r^2) \Big[ m^2 +
\fr{\ell^2}{r^2}
 -\fr{\omega^2}{1-r^2}
\Big]. \label{6eq4} \eeq Here, in contrast to the BTZ black hole,
$E=\omega^2$ is not singled out as the energy term of  the
Schr\"odinger equation because  the dS potential also includes
this term.
%This means that the wave equation approach to dS space
%induces some unclear points.

 Considering $r^*=g(r)= \int g'(r) dr$,
one finds \beq r^*= \ln r - \fr{1}{2} \ln\Big[(1+r)(1-r)\Big],~~
e^{2r^*} =\fr{r^2}{1-r^2},~~ r^2=\fr{e^{2r^*}}{1+e^{2r^*}}.
\label{7eq4} \eeq From the above
 we confirm that $r^*$ is a tortoise coordinate
such that $r^* \to -\infty (r \to 0)$, whereas $r^* \to \infty (r
\to 1)$. We can express the  potential as a function of $r^*$
explicitly  \beq V_{dS}(r^*) = \omega^2 +
\fr{e^{2r^*}}{(1+e^{2r^*})^2} \Big[ m^2 + \fr{1+ e^{2r^*}}{e^{2
r^*}} \ell^2 - (1 + e^{2r^*}) \omega^2 \Big]. \label{8eq4} \eeq
For $m^2=1,\ell^2=0,E=\omega^2=0.01$, the shape of this takes a
potential barrier ($\frown$) located at $r^*=0$. On the other
hand, for all non-zero $\ell$,  one finds the potential step
($\lnot$) with its height $\omega^2+\ell^2$ on the left-hand side
of $r^*=0$. All potentials  $V_{dS}(r^*)$ decrease exponentially
to zero as $r^*$ increases on the right-hand side. Roughly
speaking, the outline of $V_{dS}(r^*)$  shown in Fig.8 is similar
to the potential step in Fig.1. We conjecture that travelling
waves appear near the cosmological horizon of $r^*=\infty$.
\begin{figure}
\begin{center}
\epsfig{file=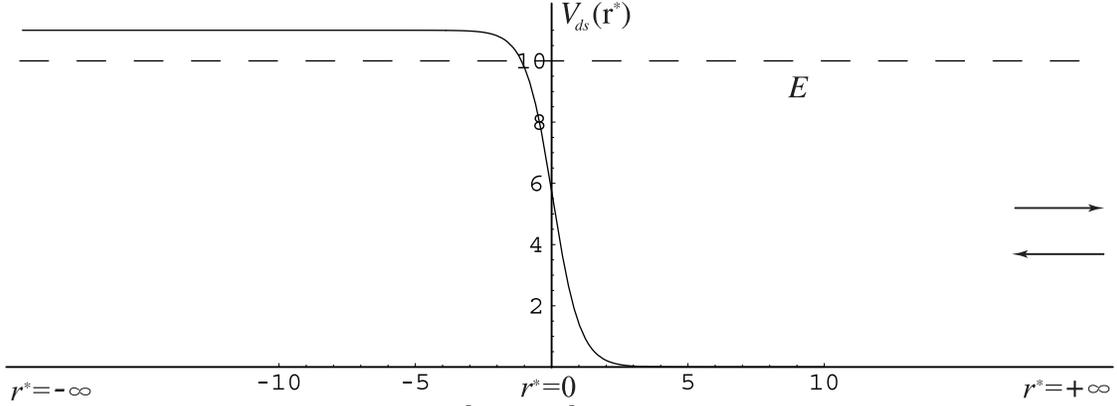,width=0.9\textwidth} \caption{The shape of
$V_{dS}(r^* )$ with $m^2 =1,~\ell^2 =1,~E=10 $. The ingoing wave
(the outgoing wave) near the cosmological horizon ($r^* =\infty$)
are indicated by $\leftarrow(\rightarrow)$ respectively. $r^*
=-\infty$ corresponds to the origin of coordinate $r=0$.}
\end{center}
\end{figure}
But near the coordinate origin of $r^*=-\infty~(r=0)$, it is not
easy to develop a travelling wave. Near $r=0~(r^*=-\infty)$ one
finds the approximate equation \beq \fr{d^2}{d r^{*2}} {\tilde
f}_{\ell,-\infty} -\ell^2 {\tilde f}_{\ell,-\infty}=0.
\label{9eq4} \eeq This  gives us a solution \beq {\tilde
f}_{\ell,-\infty}(r^*) =A_{dS} e^{\ell r^*} + B_{dS} e^{-\ell r^*}
\label{10eq4} \eeq which is equivalently rewritten by making use
of Eq.(\ref{7eq4}) as \beq {\tilde f}_{\ell,r=0}(r) =A_{dS}
r^{\ell} + \fr{B_{dS}}{ r^{\ell}}. \label{11eq4} \eeq
 In the above
two equations, the first terms correspond to  normalizable modes
at $r=0~(r^*=-\infty)$, while the second terms are
non-normalizable, singular modes. On the other hand, near the
cosmological horizon $r_c=1(r^*=\infty)$ one obtains a
differential equation which is irrespective of $\ell$ \beq
\fr{d^2}{d r^{*2}} {\tilde f}_{\ell,\infty} +\omega^2 {\tilde
f}_{\ell,\infty}=0. \label{12eq4} \eeq
 This  has a  solution \beq {\tilde f}_{\ell,\infty}(r^*) =C_{dS} ~e^{-i \omega r^*}
+ D_{dS}~ e^{i \omega r^*}. \label{13eq4} \eeq  The first/second
waves in Eq.(\ref{13eq4}) together with $e^{-i \omega t}$ imply
the ingoing $(\gets)$/outgoing $(\to)$ waves across the
cosmological horizon. These are shown in Fig. 8. This picture is
based on the observer confined in the southern diamond.

\subsection{dS flux calculation}

In order to solve equation (\ref{4eq4}) explicitly , we first
transform it into a hypergeometric equation using $z=r^2$. Here
the working space still remains unchanged as $0 \le z \le 1$
covering  the southern diamond. This equation  takes a form

\beq z(1-z){\tilde f}_{\ell}''(z) -(2z-1) {\tilde f}_{\ell}'(z) +
\fr{1}{4} \Big( \fr{\omega^2}{1-z} -\fr{\ell^2}{z} -m^2 \Big)
{\tilde f}_{\ell}(z)=0. \label{14eq4} \eeq Here one finds two
poles at $z=0,1 (r=0,1)$ and so  makes a further transformation to
cancel these by choosing a normalizable solution at $z=0(r=0)$
\beq {\tilde f}_{\ell} (z)=z^\alpha (1-z)^\beta \tilde w(z),~~
\alpha= \fr{\ell}{2},~~\beta= i\fr{\omega}{2}. \label{15eq4} \eeq
Then we obtain the  hypergeometric equation \beq z(1-z)
\tilde{w}''(z) + [c-(a+b+1)z] \tilde{w}'(z) -a b~ \tilde{w}(z) =0
\label{16eq4} \eeq where $a,b$ and $c$ are given by \beq
 a= \fr{1}{2} ( \ell + i \omega +h_+),~~ b= \fr{1}{2} ( \ell + i \omega
 +h_-),~~ c= \ell+1
 \label{17eq4}
 \eeq
 with
 \beq
 h_{\pm}= 1 \pm \sqrt{1-m^2}.
 \label{18eq4}
 \eeq
 One regular solution near $z=0$ to Eq.(\ref{14eq4}) is given
 by\cite{AS}
 \beq
 {\tilde f}_+(z)
 =A_{dS}~ z^{\ell/2}(1-z)^{i\omega/2} F(a,b,c;z)
 \label{19eq4}
 \eeq
with an unknown constant $A_{dS}$. In addition, there is the other
solution with a logarithmic singularity at $z=0$ as $ \tilde
f_-(z)=
 \tilde A_{dS}~ z^{\ell/2}(1-z)^{i \omega/2}[F(a,b,c;z) \ln z +\cdots]$.
However, both solutions have vanishing
flux at $z=0$ because the relevant part ($z^{\ell/2}$)
is not complex but real.

Now we are in a position to calculate an  outgoing flux at
$z=0(r=0,r^*=-\infty)$ which is defined as  \beq {\tilde{\cal
F}}_{out}(z=0)= 2\fr{2 \pi}{i} [{\tilde f}^*_+ z\partial_z {\tilde
f}_+ -{\tilde f}_+ z\partial_z {\tilde f}^*_+]|_{z=0}.
\label{20eq4} \eeq For any kind of real functions near $z=0(r=0)$
including $\tilde{f_{\pm}}$,
 the outgoing $(\to)$ flux is obviously given by
\beq {\tilde{\cal F}}_{out}(z=0)=0. \label{21eq4} \eeq This means
that if a wave form is real near $z=0$, one cannot find any
non-zero flux. We choose a regular  solution of $\tilde{f_+}(z)$
for further calculation. To obtain a flux at the horizon of
$z=1(r=1)$, we first  use a formula of Eq.(\ref{19eq3}) with
\beq\label{22eq4} c-a=b^*,~c-b=a^*,~c-a-b=(a+b-c)^*.\eeq Using
$1-z \approx e^{-2r^*}$ near $z=1$, one finds from
Eq.(\ref{19eq4}) the following form:

\beq {\tilde f}_{+,0\to1}\equiv {\tilde f}_{in} + {\tilde
f}_{out}=
 {\tilde H}_{\omega,\ell} ~e^{-i \omega r^*} +{\tilde H}_{-\omega,\ell} ~e^{i \omega
r^*} \label{23eq4} \eeq where \beq {\tilde H}_{-\omega,\ell}=
{\tilde H}_{\omega,\ell}^*=A_{dS}\alpha_{-\omega,\ell},~~
\alpha_{-\omega,\ell}= \fr{\Gamma(1+\ell)\Gamma(i\omega) 2^{i
\omega}} {\Gamma[(\ell +i \omega +h_+)/2)] \Gamma[(\ell +i \omega
+h_-)/2)]}. \label{24eq4} \eeq Then we match Eq.(\ref{13eq4}) with
Eq.(\ref{23eq4}) to yield $C_{dS}={\tilde H}_{\omega,\ell}$ and
$E_{dS}={\tilde H}_{-\omega,\ell}$ near the cosmological horizon.
Finally we calculate its outgoing $(\to)$ flux at
$z=1(r^*=\infty)$ as \beq {\tilde {\cal F}}_{out}(z=1)= \fr{2
\pi}{i} [{\tilde f}_{out}^*\partial_{r^*} {\tilde f}_{out}-{\tilde
f}_{out}
\partial_{r^*}
{\tilde f}_{out}^*]|_{r^*=\infty} = 4 \pi \omega A_{dS}^2
|\alpha_{-\omega,\ell}|^2. \label{25eq4} \eeq On the other hand,
the ingoing $(\gets)$ flux is given by \beq {\tilde{\cal
F}}_{in}(z=1)= \fr{2 \pi}{i} [{\tilde f}_{in}^*\partial_{r^*}
{\tilde f}_{in}-{\tilde f}_{in}
\partial_{r^*}
{\tilde f}_{in}^*]|_{r^*=\infty} =- 4 \pi \omega A_{dS}^2
|\alpha_{\omega,\ell}|^2. \label{26eq4} \eeq Thus one finds that
${\tilde{\cal F}}_{out}(z=1)+{\tilde{\cal F}}_{in}(z=1)=0$ because
of
 $|\alpha_{\omega,\ell}|^2$ =
$|\alpha_{-\omega,\ell}|^2$.

\begin{figure}[b!]
\begin{center}
\epsfig{file=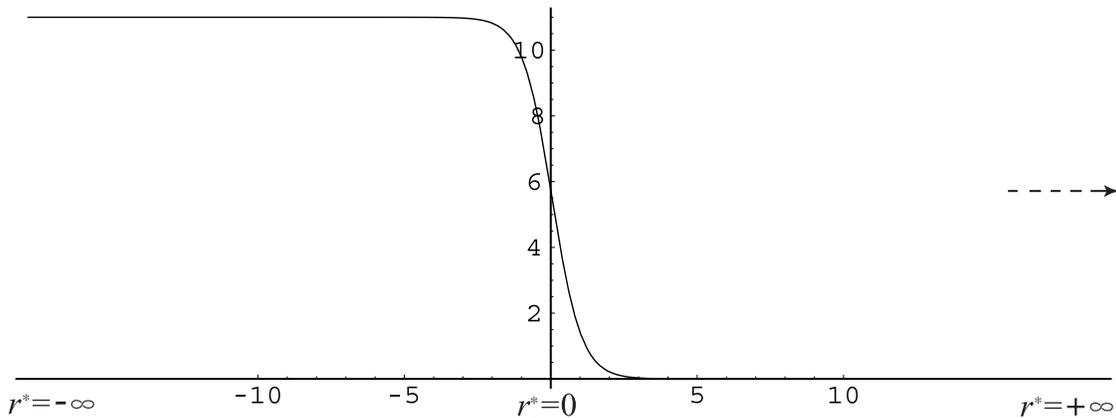,width=0.9\textwidth} \caption{The assumed
quasinormal modes ($-->$) in de Sitter space. However, these are
hard to exist in dS space.}
\end{center}
\end{figure}

\subsection {dS absorption cross section}
 Up to now we do not insert
the curvature radius $l$ of dS space. The correct absorption
coefficient can be recovered when replacing $\omega(m)$ with $
\omega l(m l)$. An absorption coefficient by the cosmological
horizon is defined formally by \beq {\tilde{\cal A}}_{dS} = \fr{
{\tilde{\cal F}}_{out}(z=1)}{{\tilde{\cal F}}_{out}(z=0)}.
\label{27eq4} \eeq  It is found that there is no absorption of a
scalar wave in  de Sitter space in a semiclassical way. This means
that de Sitter space is usually stable and in thermal equilibrium
with the scalar perturbation, unlike the AdS-black hole. The
cosmological horizon not only absorbs radiation (scalar
perturbation) but also emits that previously absorbed by itself at
the same rate, keeping the curvature radius $l$ of de Sitter space
fixed. It can be proved by the relation of $\tilde{{\cal
F}}_{out}(z=1) +\tilde{{\cal F}}_{in}(z=1)=0$ and $\tilde{{\cal
F}}_{out}(z=0)=0$. This picture coincides with the plane wave
propagation of the energy $E$ under the potential step with
$0<E<V_0$ which shows the classical picture of what goes on, as is
shown in Sec. II-A.

\subsection{dS quasinormal modes}
Because one cannot find any travelling wave at $r=0$, we assume
the definition of quasinormal modes to the flux boundary condition
in dS space : the outgoing flux ($\tilde{{\cal F}}_{out}(z=1)>0$)
at the cosmological horizon and the zero-flux ($\tilde{{\cal
F}}(z=0)=0$) at $r=0$. See Fig.9 for the assumed quasinormal
modes. Now let us check whether or not these modes exist in dS
space. One may choose $\tilde{H}_{\omega,\ell}=0$ to get the
quasinormal modes. This imposes the restriction as \beq
\label{28eq4} c-a=-n,~~c-b=-n,\eeq where $n=-1,-2,\cdots$(negative
integers). Considering Eq.(\ref{30eq3}) leads to \beqa
\label{29eq4} \omega_1=
 -2\fr{i}{l} \Big(n+\fr{1}{2}(l+1)
-\fr{\sqrt{1-(ml)^2}}{2} \Big),\\\label{30eq4} \omega_2=
-2\fr{i}{l} \Big(n+\fr{1}{2}(l+1) +\fr{\sqrt{1-(ml)^2}}{2} \Big).
\eeqa However, this condition of Eq. (\ref{28eq4}) leads through
Eq.(\ref{22eq4}) to $b^*=-n, a^*=-n$ which implies that
$\tilde{{\cal F}}_{out}(z=1)=0$ because its dependence of
$|\alpha_{-\omega,\ell}|^2$. Also their complex conjugates of
$\omega^*_1,\omega^*_2$ which comes from
$\tilde{H}_{-\omega,\ell}=0$ are not the quasinormal modes because
their outgoing fluxes are zero and $\omega_I<0$. Hence we cannot
define the quasinormal modes in dS space\footnote{Authors in
\cite{ACL} have found quasinormal modes in dS space. However,
their calculation was based on the FT and PT where are outside the
SD in Fig. 7. These regions are causally disconnected to an
observer at $r=0$ in the SD. Actually it is difficult to define
the wave propagation well in the FT and PT. In this sense their
computations are different from ours in the SD.}. This result is
consistent with the picture of stable cosmological horizon because
the presence of quasinormal frequencies implies that  the dS
scalar wave is loosing its energy continuously into the
cosmological horizon.

\subsection{dS/CFT correspondence}
We find that one cannot obtain the purely outgoing flux near the
cosmological horizon. Hence there is no  decay of a bulk scalar
into the cosmological horizon. Hence we do not obtain quasinormal
frequencies.  Because the cosmological horizon is stable, there is
no off-equilibrium configuration by the scalar perturbation. In
turn there is no perturbation of the boundary thermal states in
CFT. The thermal dS/CFT correspondence\footnote{This can be
considered as an extension of the  assumed dS/CFT correspondence
to study dS space\cite{STR}. The assumed dS/CFT correspondence can
be checked somewhere by calculation of entropy \cite{entropy} and
conserved quantities\cite{cons} in the FT or PT where is causally
disconnected to an observer in the SD. But one cannot confirm this
thermal correspondence  by making use of the wave equation
approach in the SD of dS Space.} is not established in the wave
equation approach.

\sect{summary}
We study the wave equation for  a  massive scalar
 in three-dimensional AdS-black hole and dS spaces.
 Our results are summarized in TABLE.
 For the AdS-black hole, the absorption cross section is not obtained directly from the bulk wave
 equation approach because the  space is asymptotically not flat but anti de Sitter.
 Instead, one could obtain it from the dual CFT\cite{Birm3,Birm1}.
 On the other hand, quasinormal frequencies are found and
 the AdS/CFT correspondence is established.
%\begin{center}
\begin{table}
 \caption{Comparison of the AdS-black hole with de
 Sitter space in the wave equation approach.}
 %\begin{ruledtabular}
 \begin{tabular}{lp{5.5cm}p{5cm}}
 Physical consequences   & AdS-black hole & dS space \\ \hline
 Absorption cross section   & Not found but it can be obtained from
 its dual CFT. & 0 \\
 Quasinormal modes & Well-defined and it is consistent with the
 CFT results.   & Not defined.\\
 Bulk/boundary correspondence & Thermal AdS/CFT correspondence is realized.      &
 Thermal dS/CFT correspondence is not realized.
 \end{tabular}
 %\end{ruledtabular}
 \end{table}
 %\end{center}

For dS space with the cosmological horizon, one finds that the
absorption cross section is  zero.  By analogy of the quantum
mechanics with the potential step,
 it is evident that there is no
absorption of a scalar wave   in  de Sitter space. This means that
de Sitter space is usually stable  and it is in thermal
equilibrium with the scalar perturbation. Explicitly the
cosmological horizon not only absorbs scalar waves  but also emits
those previously absorbed by itself at the same rate. As a result,
de Sitter space keeps the curvature radius $l$  fixed. This can be
proved by the flux conservation of $ \tilde {\cal F}_{out}(z=1) +
\tilde {\cal F}_{in}(z=1)=0$ and $\tilde {\cal F}(z=0)=0$. It is
identical to that in the wave propagation of the energy $E$ under
the potential step with $0<E<V_0$, which shows the classical
picture of what goes on.  This contrasts to the case of the
AdS-black hole. The quasinormal modes are not defined because one
cannot find the purely outgoing flux at the cosmological horizon.
Also the thermal dS/CFT correspondence is not realized in the wave
equation approach.

Finally we wish to mention two important results. First we clarify
that the absorption cross section  is not derived from the bulk
scalar wave propagation in the background of the AdS-black hole.
But this could be obtained from  the dual CFT through the AdS/CFT
correspondence. It implies that the shaded region of AdS-black
hole space in Fig.3 is too small to find the greybody factor.
Second, thermal properties of the cosmological horizon are
different from those of the event horizon of the AdS-black hole.
The cosmological horizon is in thermal equilibrium with the scalar
perturbation whereas the event horizon of the $J=0$ BTZ black hole
is not in equilibrium with the scalar perturbation because the
scalar waves decay into the horizon continuously.

\section*{Acknowledgement} This  was supported in part by  KOSEF,
Project No. R02-2002-000-00028-0.

\end{document}